# TEMPERATURE DEPENDENCE OF THE PROBABILITY OF "SMALL HEATING" AND SPECTRUM OF UCNs UP-SCATTERED ON THE SURFACE OF FOMBLIN OIL Y-HVAC 18/8

V.V. Nesvizhevsky[1], A.Yu. Voronin[2], A. Lambrecht[3], S. Reynaud[3], E.V. Lychagin[4], A.Yu. Muzychka[4], G.V. Nekhaev[4], A.V. Strelkov[4]

[1]*Institut Laue-Langevin, 38042, Grenoble, France*
[2]*Lebedev Institute, 119991, Moscow, Russia*
[3]*Laboratoire Kastler Brossel, UPMN-Sorbonne Universities, CNRS, ENS-PSL Research University, College de France, Campus Jussieu, 75252, Paris, France*
[4]*Joint Institute for Nuclear Research, 141980, Dubna, Russia*

**Abstract**

We performed precision measurements of the probability of "small heating" and spectrum of UCNs up-scattered on the surface of hydrogen-free oil "Fomblin Y-HVAC 18/8" as a function of temperature. The probability is well reproducible, does not depend on sample thickness and does not evolve in time. It is equal $(9.8\pm0.2)\cdot10^{-6}$ at the ambient temperature. The spectrum coincides with those measured with solid-surface and nanoparticle samples. Indirect arguments indicate that spectrum shape weakly depends on temperature. Measured experimental data can be satisfactory described both within the model of near-surface nanodroplets and the model of capillary waves.

**Introduction**

The phenomenon named "small heating" of ultracold neutrons (UCNs) was found over 15 years ago (1), (2). It consists of inelastic reflection of UCNs from surface accompanied with energy change (increase or decrease) comparable to initial UCN energy. UCNs with increased energy are called "Vaporized" UCNs (VUCNs). Measured probabilities of this process largely exceed values estimated within theoretical models considering the process of neutron reflection from bulk materials (3), (4), (5).

Small heating (and "cooling") of UCNs was measured by different experimental teams both in the reflection of UCNs from solid surfaces (1), (2), (6), (7), (8), (9), (10), and in the reflection of UCNs from liquids (1), (2), (11), (12), (9), (10). Small heating probabilities measured by different teams with the same materials differ considerably from each other. This discrepancy could be associated both with actual difference in the process probability as a function of (a) poorly controlled parameter(s) or/and with insufficient knowledge of spectral characteristics and thus insufficient precision of estimations of VUCN detection efficiency.

For solid surfaces, this contradiction was withdrawn when work (13) showed that the probability essentially changes as a function of sample history, in particular the degree of sample preceding heating. Results of a series of experimental works (7), (8), (13) and theoretical works (14), (15), (16) indicate that small heating of UCNs is due to their scattering on nanoparticles in a state of physical adsorption on surface (17). Most nanoparticles are strongly bound to surface, but some are weakly bound and thus move along the surface. UCN scattering at moving nanoparticles is the reason of their small heating. All experimental data available for solid surfaces agree well with predictions of this model.

For liquid surfaces, contradictions are still there. In general, small heating probability is measured from a certain initial UCN energy range to a range of larger energies available for spectroscopy; both ranges are specific for concrete experimental setups; correct comparison of different results requires the knowledge of both the experimental details and the efficiency of VUCN detection as a function of energy. The first observation of this phenomenon (1), (2), (6) provided a small heating probability equal $10^{-5}$ per UCN bounce from surface; an increase of UCN energy was estimated to be about equal to the initial UCN energy. Work (12) showed that UCN reflection from oil surface is accompanied both with evens of energy increase and energy decrease ("cooling"); the energy change was also estimated to be about equal to the initial UCN energy but the probability was estimated to be equal $10^{-6}$. Authors of work (9) confirmed the existence of small heating and cooling of UCNs and estimated the probability to be equal $3 \cdot 10^{-6}$; in work (10) the probability was estimated to be equal $5 \cdot 10^{-6}$. As all these measurements were carried out at the same (ambient) temperature and VUCN spectra were measured or at least simulated, this scattering of results exceeds estimated experimental accuracy.

In the present work, we study small heating of UCNs on the surface of hydrogen-free oil Fomblin®. In spite of a number of related publications, small heating of UCNs on Fomblin has not been investigated in detail. In particular, nobody has measured the spectrum of up-scattered neutrons. A physical origin of this phenomenon on liquid surfaces could differ from that on solid surfaces. Thus works (18), (19) consider the interaction of UCNs with surface capillary waves as a reason for small heating. Investigation of small heating on liquid surfaces is interesting also in view of its contribution to systematic errors in measurements of the neutron lifetime (20). Recent work (21) revises results of some experiments involving UCN storage in Fomblin traps from this point of view.

Hydrogen-free oils of various types are used in UCN experiments. They could be subdivided in two groups: low-temperature and high-temperature oils. Solidification temperatures of low-temperature oils are lower, saturation vapor pressures are higher; such oils can operate at nitrogen temperature and provide lowest UCN loss coefficients (22). High-temperature oils are widely used due to more convenient operation and the absence of mass transfer at ambient temperatures. In the present work we study high-temperature oil named Fomblin Y-HVAC 18/8 (23); it was used in neutron lifetime experiments (24), (25), (26), (27), (28), (29).

**Experimental setup and measuring procedure**

The experiments is carried out in the Big Gravitational Spectrometer (BGS) built specially to investigate small heating of UCNs. Detailed description of the spectrometer and measuring procedure is found in work (13). The spectrometer is a vertical cylindrical UCN trap with an absorber in the upper part. Lower part of the trap is divided in two parts, internal and external, using a gravitational barrier. A scheme of principle and a photo of the internal part are shown in Fig. 1. Sample (1) is placed to the spectrometer bottom inside the internal part, which is defined by a cylinder (2) with the diameter 40 cm and the height 35 cm. The cylinder is the gravitational barrier for those UCNs, which filled in the internal volume through an input neutron guide and then are trapped inside with a valve (3). UCN flux density at the spectrometer bottom is monitored using a monitor detector (4). An exit from the storage volume to the monitor detector is closed with a valve, which is a cover that can move up-down. A calibrated hole (8) with the diameter 5 mm is made in the valve. Samples can be inserted into the central part of the storage volume and extracted from it using a "lift" (not shown in the figure), which is built of two horizontal disks connected to each other with three vertical pillars. A sample is placed onto the lower disk; its upper surface is a part of the storage volume bottom when the "lift" is moved up. When the "lift" is moved down, a sample is found in a service volume; in this case, the upper surface of the upper disk is a part of the storage volume bottom, instead of the upper surface of the lower disk. The "lift" allows measuring with/without sample without breaking vacuum.

Spectrum of UCNs stored in the spectrometer is shaped from above with an absorber (5) installed at a certain height $H_{abs}$ (below we will indicate UCN energy in units of height of maximum raising in the Earth's gravitational field). Neutrons with energy sufficiently high for raising up to a height larger than $H_{abs}$ are being lost in the absorber with a certain characteristic time. The absorber consists of two parts; the central part can be inserted inside the internal part of the storage volume.

A detector (6) is installed outside the gravitational barrier (2); thus it can count only neutrons with energy higher than the barrier ($E_{bound}$). If the absorber is moved to its upper position ($H_{abs} >$ $E_{bound}$), neutrons with energy higher than $E_{bound}$ can escape above the gravitational barrier and find the detector (6) through an open input valve (7).

A device for inserting an additional neutron absorber is installed in the bottom of the external part of the storage volume (not shown in the figure). This absorber assists eliminating neutrons in the external storage volume more efficiently. This absorber is produced of titanium; its total surface area is 720 cm². 

Compared to the setup described in ref. (13), we did the following modifications in the spectrometer construction. The gravitational barrier height is decreased from 50 cm to 35 cm; thus the range of observable energy transfers increased towards higher transfer values. The titanium absorber (5) is replaced with a polyethylene absorber with developed surface; this modification allows sharply shape initial UCN spectrum and thus decrease a «dead-zone» – a range of small energy transfers that cannot be observed. The input neutron guide is inserted into the storage volume by 25 cm above its bottom; thus the energy "mono-line" in the initial UCN spectrum is quite narrow.

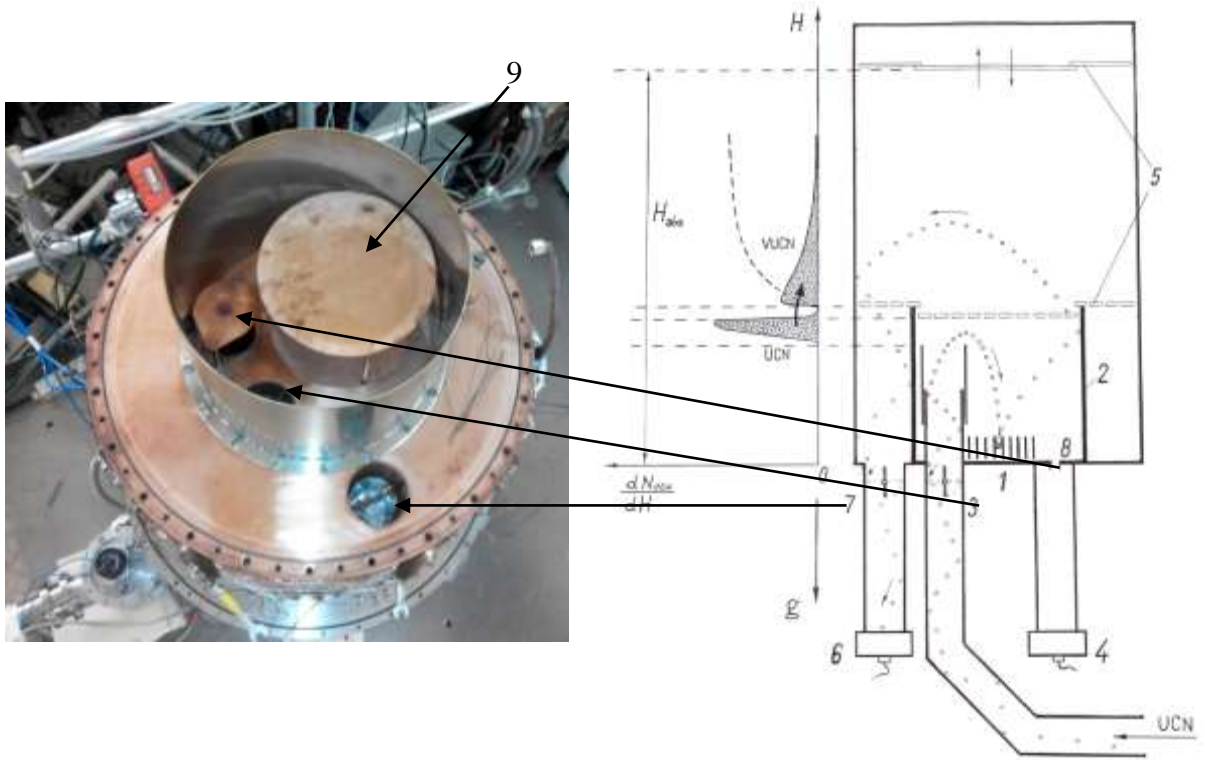

*Fig. 1. A scheme of the setup. 1 – sample, 2 – gravitational barrier, 3 – input valve, 4 – monitor, 5 – absorber, 6 – detector, 7 – exit valve, 8 – calibrated hole.*

The procedure to measure small heating and spectrum of heated neutrons consists of the following stages: «filling», «cleaning», «effect measurement», «spectrometer emptying». Fig. 2 shows typical evolution of the detector (6) and monitor (4) count rates during these stages in measurements with a sample. During all these stages, the exit valve in front of the detector is open, the valve in front of the monitor detector is closed.

During filling in the spectrometer (0-60 sec), the input valve is open, the absorber is found in its lower position $H_{abs} = H_{min}$ at the height 32.5 cm. Some UCNs from initial spectrum with energy higher than the gravitational barrier escape through a slit between the absorber and the walls of internal storage volume, jump over the gravitational barrier and find the detector; therefore the count rate is high.

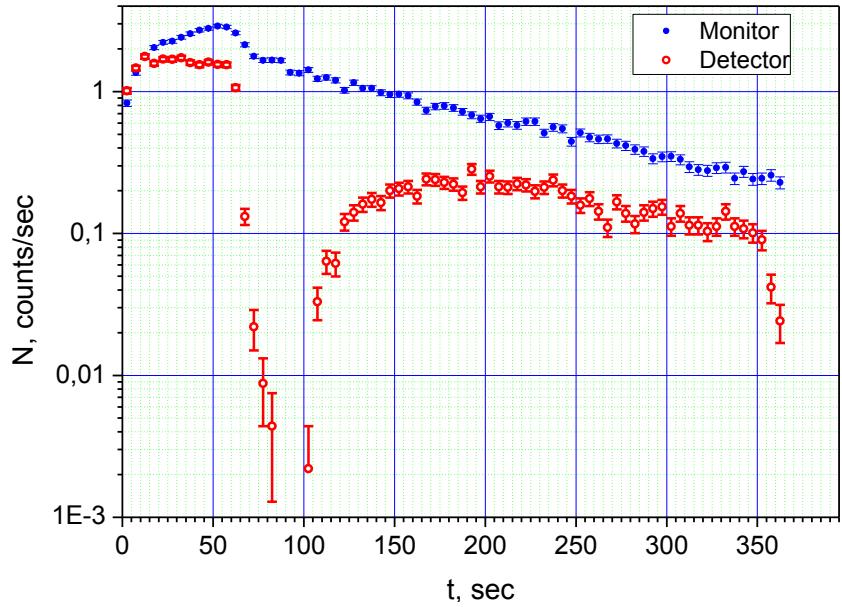

*Fig. 2 Count rate in the detector (open points) and the monitor (solid points) as a function of time in measurements with a sample (explanations in the text). The sample surface area is 0.74 m²; the measurement is performed at the ambient temperature; the maximum absorber height is 140 cm.*

After closing the input valve (60-th sec) the cleaning starts. Neutrons with energy higher than the absorber height are promptly lost in it or escape from the external volume to the detector. As a result, count rate in the detector sharply decreases down to the background value. Only neutrons with energy lower than $H_{min}$ survive in the central part of the storage volume. The duration of cleaning $\Delta t_{clean}$ is 40 sec; it is chosen so that neutrons with energy larger than $H_{min}$, which survived in the internal storage volume, would not affect the result. Procedures of choosing values $\Delta t_{clean}$ and $H_{min}$ and measuring UCN spectra are described in Appendixes 1 and 3.

At the end of cleaning (110-th sec), the absorber is lifted to the upper height $H_{abs} = H_{max}$, and the effect measurement starts. Raising of the absorber does not affect neutrons with energy lower than $H_{min}$ and thus does not affect the monitor count rate, while the detector count rate increases and after some time becomes proportional to the flux density of UCNs trapped in the spectrometer with the gravitational barrier. This dependence is explained by permanent production of neutrons (VUCNs) in the storage volume with energy higher than the gravitational barrier.

If an UCN gains energy larger than $H_{max}$ due to inelastic scattering on spectrometer or sample surfaces than a probability of finding the detector is suppressed by its loss in the absorber. Thus the detector counts mainly UCNs with energy smaller than $H_{max}$, and one can evaluate the integral VUCN spectrum by means of comparing results measured with different $H_{max}$. In order to calculate spectrum using measured numbers of neutrons counted in the detector as a function of the absorber height (effect measurement) one should take into account that the efficiency of neutron detections depends on their energy. This is due to the fact that both the probability of neutron escape to the detector and the probability of its loss in the spectrometer change as a function of energy; the later only changes as a function of sample in the spectrometer. However, these values can be measured and thus the efficiency of counting VUCNs as a function of energy can be measured experimentally. Fig. 3 shows the probability of VUCN detection as a function of energy for different absorber heights. A procedure of measurement of the neutron detection efficiency as a function of energy is described in Appendix 2.

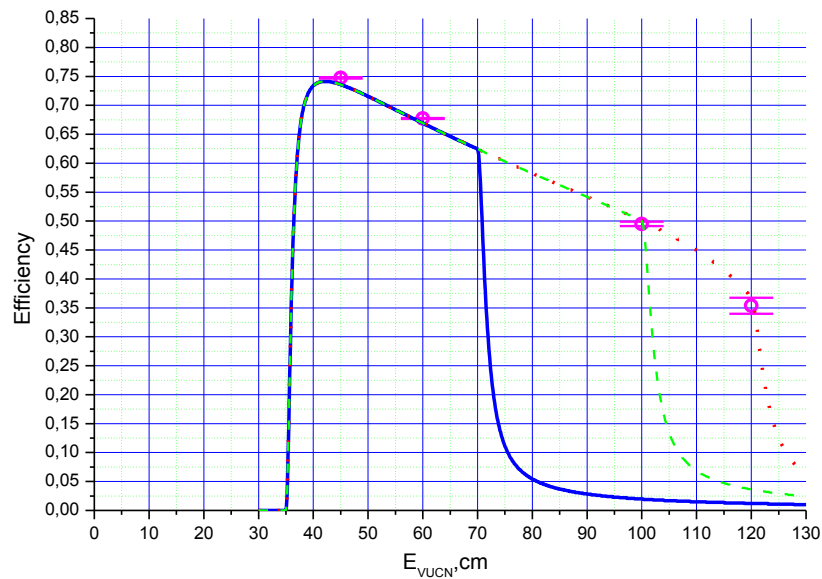

*Fig. 3. VUCN detection efficiency as a function of energy. Points correspond to experimental values measured with a sample. The sample surface area equals 0.74 m². Curves indicate calculations for different absorber heights: 70 cm (solid curve), 100 cm (dashed curve), and 120 cm (dotted curve).*

At 350-th sec, the absorber is moved to the lower height and the detector count rate decreases again to the background value.

Statistics is accumulated by periodically repeating the measuring procedure.

**Samples**

In order to study small heating of UCNs on the surface of Fomblin oil we produced samples of two types: thin layers and thick layers of oil. Thin layer means a thickness of about 1 μm; thick layer means a thickness of much larger than 1 μm, up to a few mm (such thicknesses were applied in ref. (29); a thin layer at side trap walls; a thick layer at the bottom).

Thin layers are applied to a stainless steel foil. The foil was dipped into a bath with oil and then taken off. Then oil flew down from the vertically installed foil for over 24 hours in a clean room for preventing dusting of the surface. The sample surface area was equal 0.74 m$^2$.

To produce a thick sample, we filled in flat "plates" with layers of oil with the thickness 2-3 mm; the plates were coated with a thin oil layer from all other directions. In order to increase the sample area up to 0.41 m$^2$, we installed five "plates" above each other. Figs. 4, 5 and 6 show images of sample of the two types.

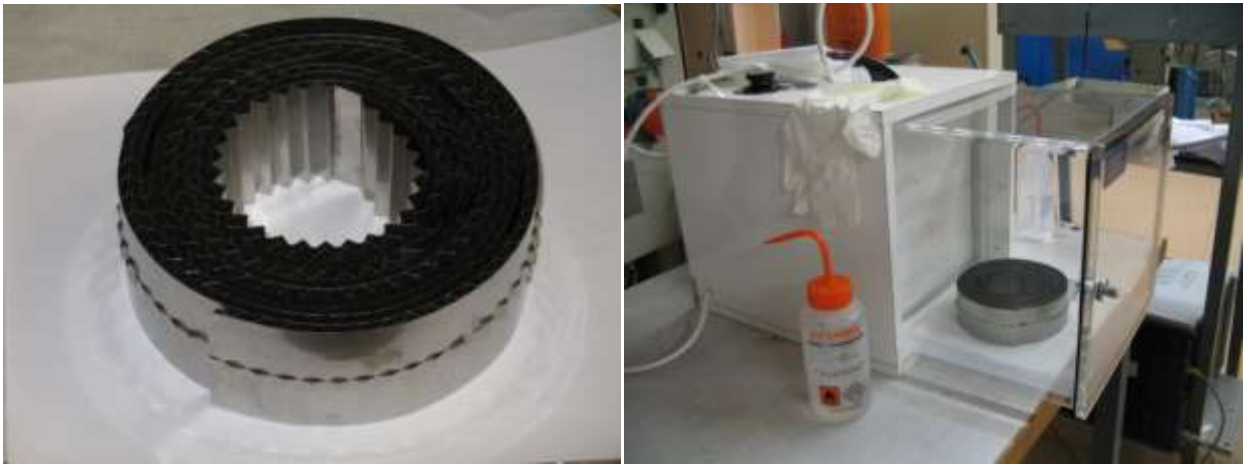

*Fig. 4. A stainless steel foil and the foil coated with oil in a portable "clean room".*

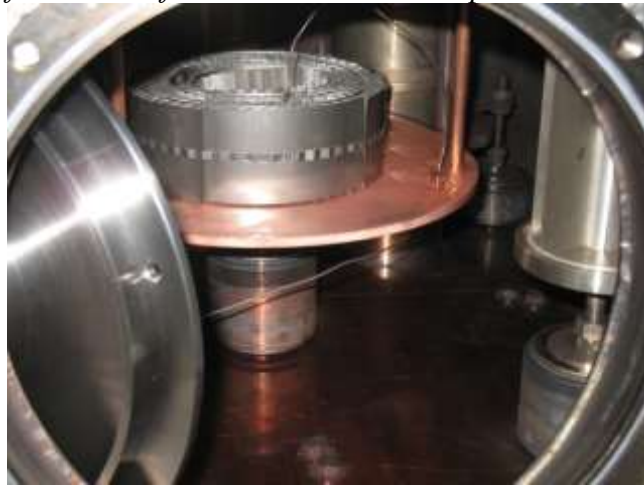

*Fig. 5. A thin-layer oil sample in the service volume of the spectrometer.*

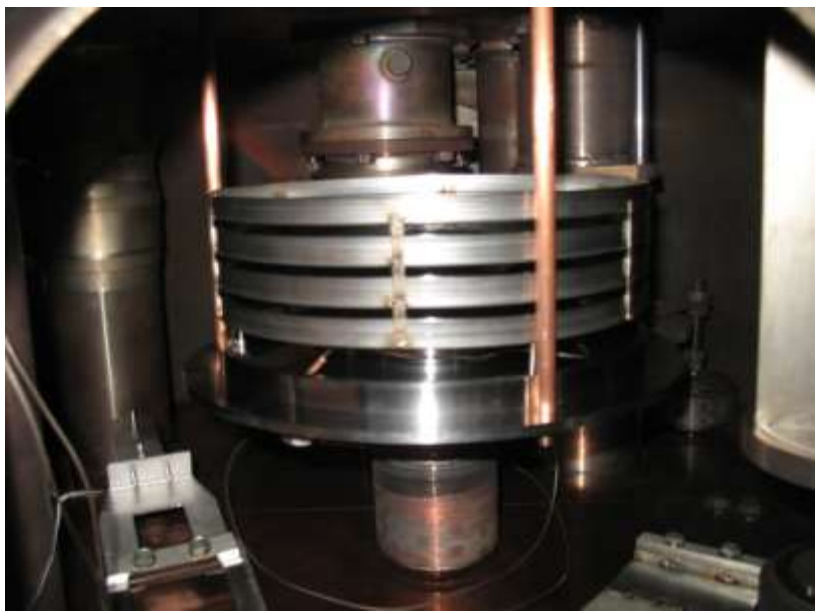

*Fig. 6. "Plates" with oil in the service volume of the spectrometer.*

We use Fomblin Y-HVAC 18/8 oil from Ausimont (now SolvaySolexis) company to produce thin samples (30), (31), (32). Thick samples are produced both from Ausimont-company oil and from oil previously used in ref. (33).

The pressure of Fomblin Y-HVAC 18/8 oil vapor pressure at the ambient temperature is as low as $\sim 2 \cdot 10^{-8}$ mbar that favor its utilization at this temperature and below, without being afraid of mass transfer to the spectrometer surfaces.

In addition to oil samples, we used samples of diamond nanopowder (nano-diamond produced in accordance with ТУ 2-037-677-94ФГУП "РФЯЦ–ВНИИТФ", Snezhinsk) and sapphire nanopowder (34). The characteristic diameter of diamond nano-crystals is 20 nm. Nano-powder samples are in the shape of layer with the thickness of up to 1 mm on copper or single-crystal sapphire surface. A nano-diamond sample is shown in Fig. 7.

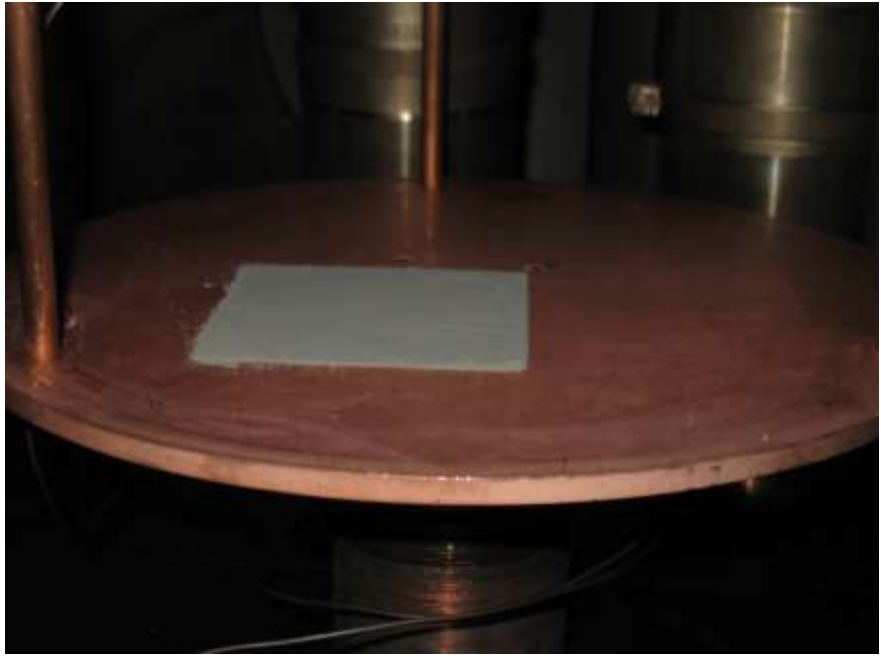

*Fig. 7. Nano-diamond powder on the spectrometer surface.*

**Results of measurements**

*a) UCN spectrum*

The initial UCN spectrum in the spectrometer is shown in Fig. 8.

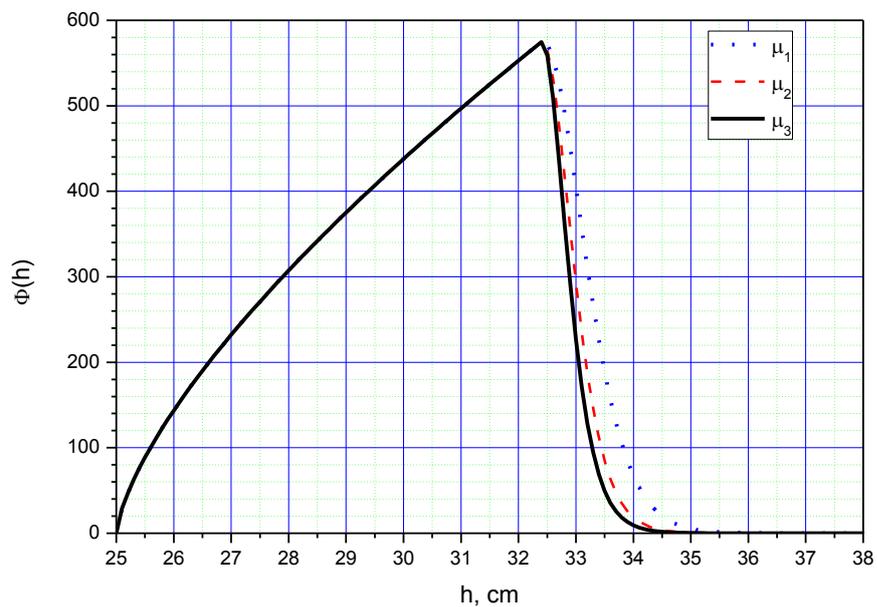

*Fig. 8. Differential UCN spectrum in the spectrometer. Fractions of spectra on right are calculated within different absorber models (see Appendix 1).*

Procedures of measurement and spectrum calculation are described in Appendix 3.

As clear from Fig. 8, the main fraction of UCNs is found in the energy range $30.5 \pm 2.5$ cm.

*b) Probability of small heating on Fomblin*

All parameters of measurements are nicely reproducible for all Fomblin samples; they do not evolve in time. Results for thin and thick samples do not differ from each other within uncertainties. The probability of small heating at the ambient temperature is the same for all samples. The probability from the given initial energy range to the range of final energies up to 120 cm (see Fig. 3, the absorber height is 120 cm) is equal (9.8±0.2)·10⁻⁶ per collision with surface. This value is estimated using equation (1):

$$P_+ = \frac{N_{det}}{N_{mon} \cdot \varepsilon} \frac{S_{mon}}{S_{sample}}, \qquad (1),$$

where $P_+$ is the small heating probability, $N_{det}$ and $N_{mon}$ are the integral count rates in the detector and the monitor during the effect measurement, $S_{mon}$ and $S_{sample}$ are the area of monitor hole and the effective sample area accordingly, $\varepsilon \approx 0.7$ is the efficiency of VUCN detection taking into account VUCN detection as a function of the absorber height.

In contrast to that, results of measurements with nano-powders evolve in time. For thick powder layers, the dominant tendency is the increasing of losses and the decreasing of small heating probability.

*c) Small heating as a function of Fomblin temperature*

The probability of small heating is shown in Fig. 9 as a function of temperature for a thin Fomblin layer; it is given in values of the ratio $N_{det}/N_{mon}$ in expression (1). Note that the probability of VUCN detection is defined mainly by VUCN losses in the spectrometer walls; it is virtually independent from Fomblin sample (losses on samples are significantly smaller than losses on spectrometer walls) and temperature in the studied range.

Black point in Fig. 9 corresponds to measurements at the ambient temperature; blue points indicate measurements during cooling; and red points show results measured during heating.

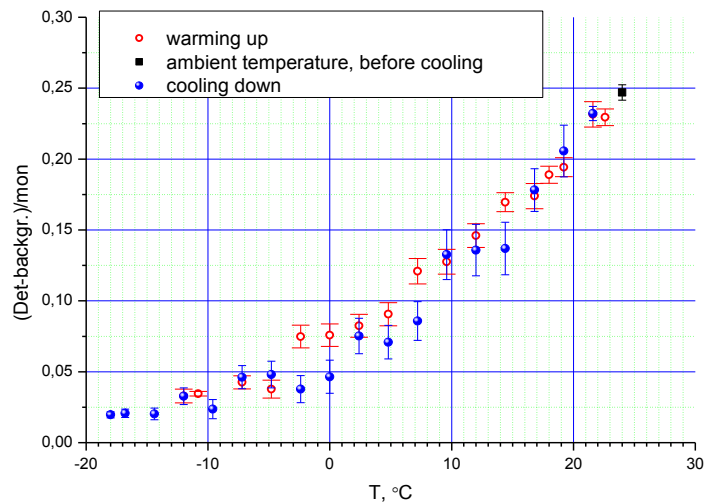

*Fig. 9. Small heating as a function of temperature; a thin Fomblin layer; explanations in the text.*

Fig. 9 illustrates measurements of small heating as a function of temperature.

*d) VUCN energy distribution*

Fig. 10 presents the number of detected up-scattered neutrons as a function of the absorber height during the effect measurement; neutrons are up-scattered on different Fomblin samples and on spectrometer walls.

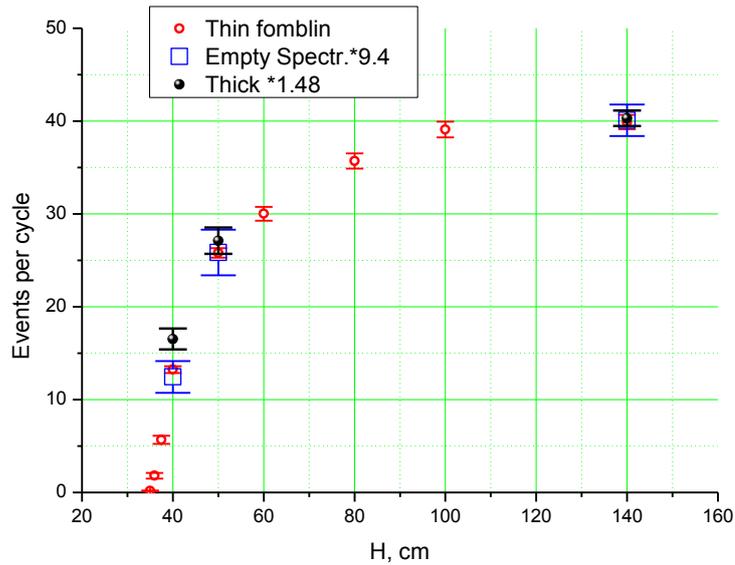

*Fig. 10. The number of detected up-scattered neutrons as a function of the absorber height; neutrons are upscattered on different samples during the effect measurement. Open circles correspond to measurements with thin Fomblin layers; open squares indicate results obtained with empty spectrometer and multiplied with a factor 9.4; solid points show results measured with thick Fomblin layers and multiplied with a factor 1.48.*

Analogous measurements were carried out with nanoparticles of all types. Fig. 11 compares averaged numbers of detected UCNs as a function of the absorber height measured during the effect measurement for all Fomblin and nanoparticles samples. Curves are presented in relative units; they are normalized to maximum values.

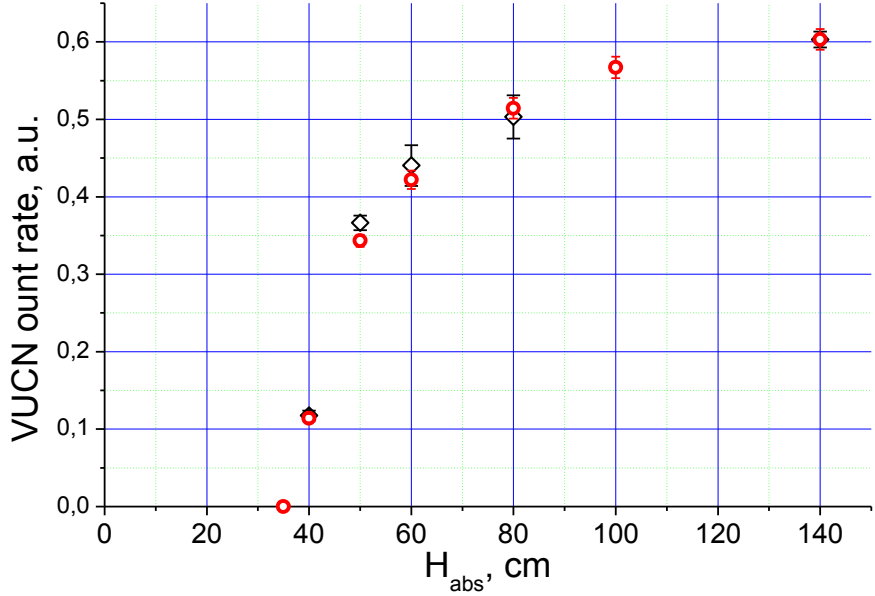

*Fig. 11. VUCN count rate as a function of the absorber height measured during the effect measurement. Circles show results measured with various solid nanoparticles: diamond, sapphire, copper; all results coincide within statistical uncertainty. Rhombuses indicate data measured with Fomblin. Curves are presented in relative units; they are normalized to maximum values.*

**Discussion of results**

As mentioned in the introduction, we are aware of two hypotheses, which can describe the phenomenon under investigation: the hypothesis of inelastic UCN scattering on near-surface nanodroplets (16) and the hypothesis of inelastic UCN scattering on surface capillary waves (18). While temperature and spectral dependencies for small heating on capillary waves are considered in ref. (19) in detail, the nanodroplets hypothesis has not been so worked out. One could assume that the probability of small heating depends on the density of droplets of a certain size in the vicinity of the surface; in a turn, the density is expected to be related to the saturation vapor pressure with a simple dependence.

We did not found any data from Fomblin producer on the oil saturation vapor pressure below the ambient temperature. A formula $og(p) = A - B/T$, where $A$ and $B$ are constants for the vapor pressure below 1 tor as proposed in ref. (35). Coefficient $B$ is related to the specific evaporation heat. Coefficients $A = 3.9$ and $B = 2763$ proposed in ref. (35) do not describe satisfactory the data of Fomblin producer for the saturation vapor pressure (23). However, the same analytical formula with other coefficients $log(p) = 11.103 - 5509/T$ describes these data significantly better. The value $B = 5509$ is evaluated using values $\Delta H^V = 9\ cal/g$ and the molecular mass 2800. The value $A = 11.103$ is fitted so to obtain the known pressure value at the temperature $T = 293\ K$. Square root of saturation vapor pressure is shown in Fig. 12 as a function of temperature and compared to the experimental data.

The probability of small UCN heating on capillary waves is calculated in accordance with formulas in ref. (19) from the energy mono-line 33 neV to the energy range 38-106 neV and also shown in Fig. 12 as a function of temperature.

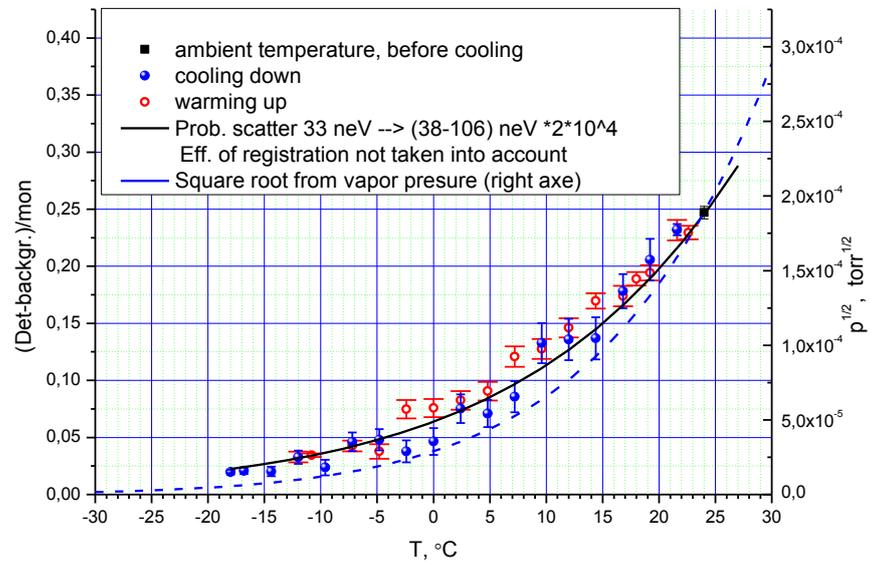

*Fig. 12. The small heating probability as a function of temperature (as in Fig. 9) compared with calculations within the hypothesis of capillary waves (black solid curve) as well as with square root of the Fomblin saturation vapor pressure (blue dashed curve, scale on right).*

Fig. 12 illustrates that formulas in ref. (19) describe well both the absolute value (the calculated small heating probability equals $1.2 \cdot 10^{-5}$ at the temperature $24^{O}$ C) and the temperature dependence. To remind, the measured probability equals $(9,8 \pm 0.2) \cdot 10^{-6}$ per collision.

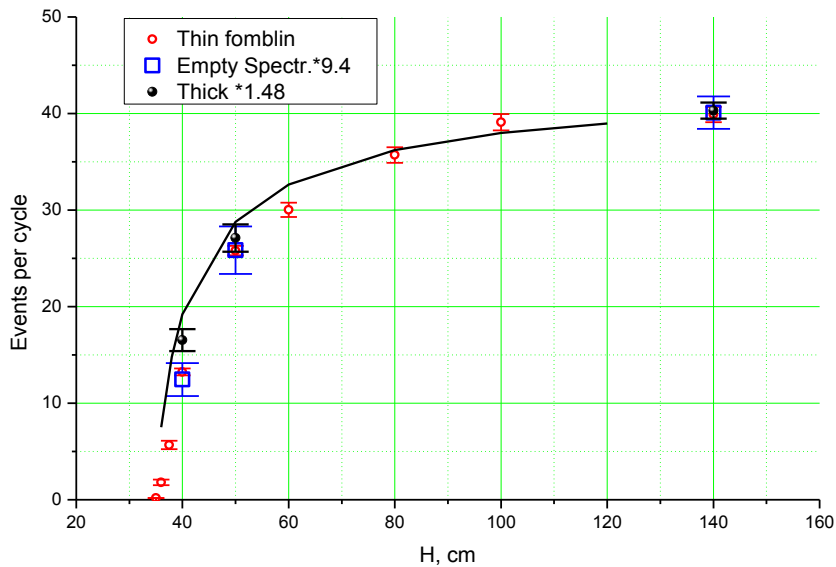

*Fig. 13. Experimental data from Fig. 10 are compared with estimations within the model of capillary waves (solid line) (19) taking into account the spectrum of UCNs in the*

*spectrometer and the VUCN detection efficiency. Count rates are renormalized to coincide at infinite height.*

VUCN count rate during the effect measurement calculated within this model is shown in Fig. 13 as a function of the absorber height; it is compared to the experimental data presented in Fig. 10. The difference between calculated and measured results is significant, in particular at small heights.

It is clear from Figs. 10 and 11 that observed VUCN count rates, as a function of the absorber height, are surprisingly very similar to each other for all solid and sample samples. This universality is natural in the model of inelastic UCN scattering on nanoparticles. Indeed, the spectrometer is selectively sensitive to some nanoparticle mass; the mass value can be calculated as a function of the initial UCN distribution and the energy window of spectrometer sensitivity. For illustration we calculated integral VUCN spectra and corresponding VUCN count rates for extreme assumptions about nanoparticle size distribution as a function of the absorber height taking into account measured spectrometer sensitivity, Fig. 14. Here, a simplified model assumes UCN scattering on individual spherical uniform nanoparticles; we neglect rotation, shape distribution, non-uniformity, interference with surface.

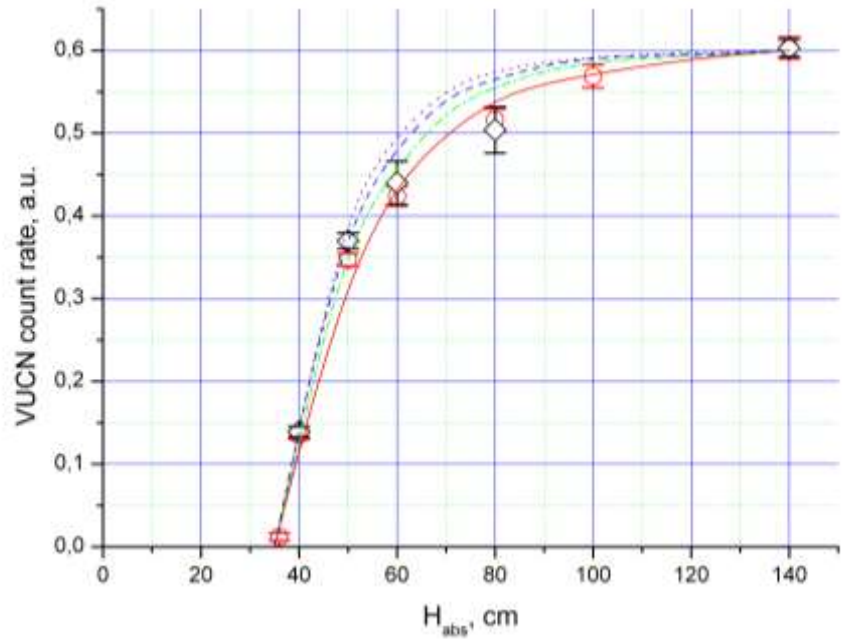

*Fig. 14. VUCN count rate as a function of energy in units of UCN raising in the Earth's gravitational field, in cm (corresponds to Fig. 11). Circles show results obtained with various solid nanoparticles: diamond, sapphire, copper; all these results coincide within statistical accuracy. Rhombuses indicate data measured with Fomblin oil; they agree with the data for solid nanoparticles. Four curves correspond to model calculations of VUCN spectra with Fomblin oil for four hypotheses on nanodroplets size distribution: $(R/R_0)^{-l}$, where index $l = 1, 2, 3, 4$ runs up down. Curves and data are normalized so to provide equal values at the infinite height.*

**Conclusion**

We performed precision measurements of the probability of small heating and spectrum of UCNs up-scattered on the surface of hydrogen-free oil Fomblin Y-HVAC 18/8 as a function of temperature. The probability is well reproducible, does not depend on sample thickness and does not evolve in time. It is equal $(9.8\pm0.2)\cdot10^{-6}$ at the ambient temperature, in agreement with the first experiment (1). The spectrum coincides with those measured with solid-surface and nanoparticle samples.

Measured results indicate that the hypothesis that UCN inelastic scattering on surface waves is the reason of small UCN heating on Fomblin, agrees to experimental data. This conclusion is based on a calculation of small heating as a function of temperature within the analytical model described in ref. (19). This is in contrast to the conclusion in our work (16), which was based on calculations illustrated in Fig. 13 in work (19); apparently the caption to Fig. 13 in work (19) is not correct. A deviation of the experimental data from the model is observed, in particular at small energy transfers.

The hypothesis of nanodroplets stays attractive, in particular due to natural interpretation of the observed universality of VUCN count rates as a function of the absorber height, as well due to the coincidence of predicted and observed spectra. However, in order to prove or rule out this hypothesis, one should develop a microscopic theoretical model of nanodroplets formation in the surface vicinity. After comparing the small heating probability and the oil vapor pressure as a function of temperature, we conclude that the probability of nanodroplets formation should be approximately proportional to square root of vapor pressure; otherwise the model would contradict the data.

The measured temperature dependence of small heating probability is analogous to that measured in ref. (10). On the other hand, the range of final energies in our setup is much closer to the initial UCN energy than that in ref. (10). Thus we conclude that apparently VUCN spectrum does not change significantly with temperature.

In neutron lifetime experiments with such Fomblin oil coatings, evadingly, UCN traps should be cooled down; cooling down to the temperature -20°C decreases the small heating probability by a factor of over 10. It will be interesting to measure the probability of small heating on oils with large molecules with correspondingly larger viscosity and lower vapor pressure. The probability of small heating on such oils might appear to be significantly lower even at the ambient temperature.

The authors are grateful to S.N. Chernyavskiy for useful discussions, also to P. Geltenbort, T. Brener, and A. Elaazzouzi for help in the experiment preparation..

**Appendix 1. Cleaning of initial spectrum**

Purposes of the "cleaning" stage are, on one hand, to shape the initial neutron spectrum so that its upper cut off is ultimately close to the gravitational barrier height, on another hand, to avoid systematic errors associated with eventual contributions of residual neutrons in the initial spectrum with energies above the gravitational barrier.

We can evaluate experimentally the cleaning time needed for shaping the initial spectrum. One method consists of measuring the effect as a function of cleaning time. Another method is to fix a cleaning time (we set $\Delta t_{clean} = 45\ sec$) and measure the effect as a function of absorber height $H_{min}$. A description of these procedures is found in ref. (13) (description of Fig. 8). Assume that we fix a certain value of cleaning time and are setting the upper cut off of the initial UCN spectrum closer and closer to the gravitational barrier. Then the measured effect will be increasing smoothly due to increasing of the total number of UCNs in the spectrometer and decreasing the energy dead-zone in the final energy spectrum. At a certain absorber height, further approaching of the spectrum cut off to the gravitational barrier will be accompanied with sharp increasing of the measured effect; it is associated with a systematic effect: neutrons from the initial spectrum with energy higher than the gravitational barrier "survive" the cleaning procedure and find the detector during counting VUCNs, Fig. 1.4. The absorber height corresponding to the transition in the derivative of the measured affect as a function of the absorber height depends on absorber parameters.

A comparison of various methods of shaping of the upper cut off in the initial spectrum was carried out in ref. (36). In present measurements we used an absorber installed in the top part of the storage volume produced from polyethylene with developed surface (see Fig. 1.1). Development of the surface increases the number of collisions of UCNs with absorber. Polyethylene absorber, and thus the spectrometer, cannot be heated up, however it provides maximum rate and sharpness of UCN spectrum shaping from above.

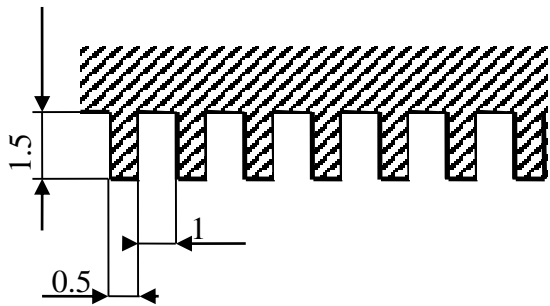

*Рис. 1.1. The absorber profile; sizes are given in mm.*

The absorber height is calibrated with a ruler. The accuracy of setting the bottom surface of the absorber in these measurements was better than 2 mm.

Losses of neutrons of interest in the absorber can be described in a simple approximation of isotropic UCN flux at the absorber height.

The measured density of polyethylene is 0.86 g/cm$^3$; the corresponding estimation of the optical potential is -8.06 neV. Here we remind main formulas for absorber with negative optical potential installed above storage volume. With the normal-to-absorber-surface component of neutron velocity $v_\perp$ and the critical velocity $v_{lim}$ corresponding to the absorber negative optical potential $-H_{lim}$, the coefficient of neutron loss per collision can be readily calculated:

$$\mu(y_\perp) = \frac{4 y_\perp \sqrt{y_\perp^2+1}}{\left(y_\perp + \sqrt{y_\perp^2+1}\right)^2} \quad (1.1)$$

where $y_\perp = v_\perp/v_{lim}$. Averaging of this coefficient over isotropic neutron distribution at the surface will provide:

$$\overline{\mu(y)} = \frac{8y}{3}\left((y^2+1)\sqrt{y^2+1} - y\left(y^2+\frac{3}{2}\right)\right), \quad (1.2)$$

here $= v/v_{lim}$, and $v$ is the velocity of incident neutron.

The rate of neutron losses in the absorber is defined by the ratio of the flux of neutrons leaving volume through absorber to the flux total flux of neutrons:

$$\frac{1}{\tau_{clean}(v)} = \frac{\frac{1}{4}\int_S n(v) v \overline{\mu(v)} dS}{\int_V n(v) dV}, \quad (1.3)$$

where $S$ is the absorber surface, $V$ is the volume available for neutrons with velocity $v$, and $n(v)$ is the neutron density. We take into account that the neutron density is proportional to velocity $n(v) \sim v = \sqrt{v_0^2 - 2gh}$, where $v_0$ is the neutron velocity at the storage volume bottom, $g$ is the gravitational acceleration, $h$ is the height above the volume bottom. If absorber is a "cover" of the storage volume with vertical walls, then for neutrons with energy higher by $\Delta h$ than the absorber height $H_{abs}$ we get:

$$\frac{1}{\tau_{clean}(\Delta h, H_{abs})} = \frac{3}{8} \frac{\sqrt{2g}\Delta h \overline{\mu\left(\frac{\Delta h}{H_{lim}}\right)}}{(H_{abs}+\Delta h)^{3/2} - (\Delta h)^{3/2}}. \quad (1.4)$$

If $\Delta h$ is small then the cleaning rate is large due to two factors: increase in the probability of reflection from absorber and decrease in the neutron flux at absorber. The probability of reflection can be lower if the absorber surface is developed. Then a neutron found in the vicinity of absorber can hit it a few times. The profile of our absorber is shown in Fig. 1.1.

The surface of our absorber was increased by a factor of 3 compared to the flat surface. As this increase places its role only for neutrons with small height exceeds $\Delta h$ (the loss coefficient for them differs significantly from 1), we will assume that its area is not affected; instead we modify expression (1.2) in the following way. Let $i$ be the number of collisions of neutron with absorber surface per one approach to the absorber surface. Then:

$$\mu_i(y) = \mu_{i-1}(y) + (1 - \mu_{i-1}(y))\mu(y) \text{ for } i > 1, \text{ and } \mu_1(y) = \mu(y). \tag{1.5}$$

Figs. 1.2 and 1.3 present loss coefficients $\mu_i(y)$ and characteristic time of cleaning $\tau_{clean}(\Delta h, H_{abs})$ as a function of $H$ and $\Delta h$ respectively.

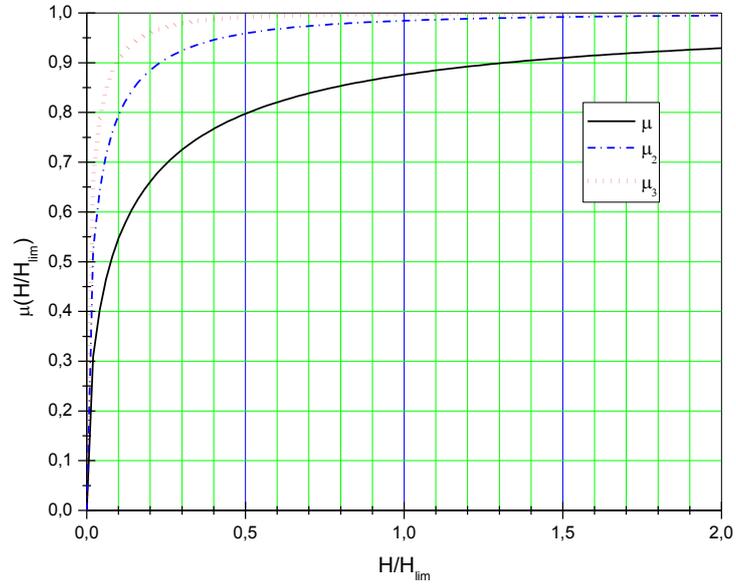

*Fig. 1.2. The coefficient of loss in polyethylene absorber as a function of neutron energy for neutrons with different numbers of neutron collisions with absorber per one approach.*

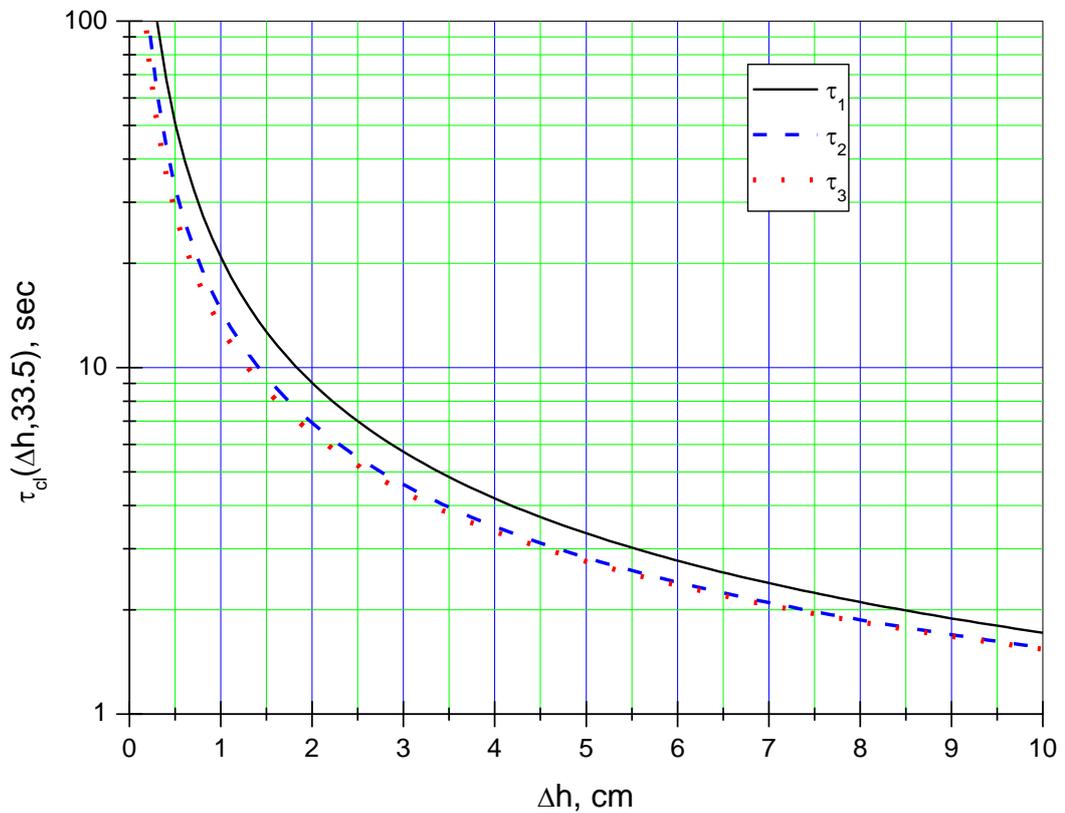

*Fig. 1.3. The characteristic time of cleaning in absorber installed at the height 33.5 cm as a function of neutron energy exceed above the absorber for different loss coefficients $\mu_i$ ($i = 1, 2, 3$).*

Fig. 1.4 illustrates measured neutron count rate during "effect measurement" in comparison with model calculations, which takes into account UCN spectrum in the spectrometer, probability of VUCN detection (see Appendix 2), and different coefficients of losses in absorber corresponding to different values $i = 1, 2, 3$ in equation (1.5).

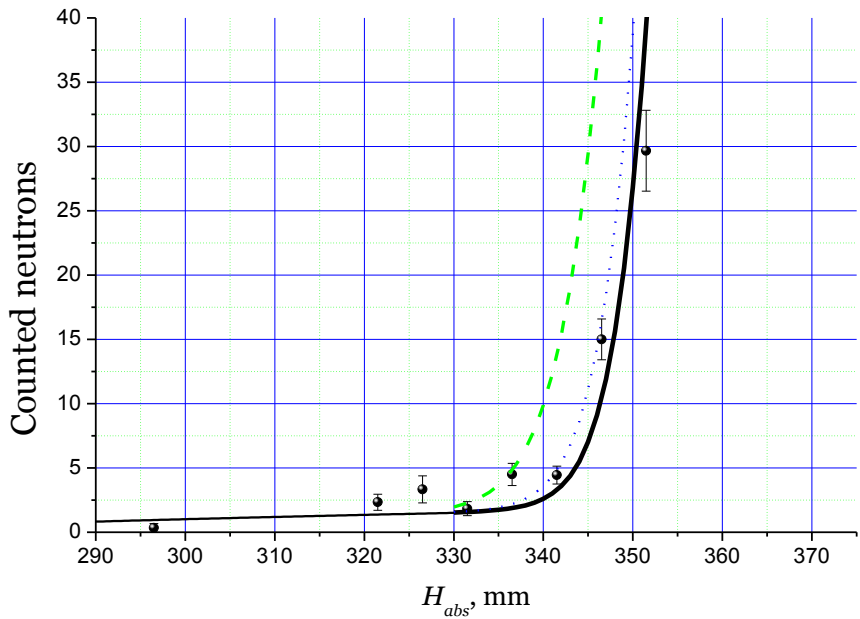

*Fig. 1.4. Observed effect as a function of the absorber height during UCN spectrum shaping. Points correspond to experimental data with empty spectrometer, curves are results of calculations with different $\mu_i$: $i = 1$ – dashed line, $i = 2$ – dotted line, $i = 3$ – solid line.*

First, Fig. 1.4 indicates that the absorber model with loss coefficient (1.5) with $i = 3$ is closest to the data; we used it for describing the efficiency of VUCN detection. Second, with a fixed cleaning time, low absorber efficiency can be compensated with decreasing its height; this modification slightly increases the energy "dead zone". As introduction of samples into the spectrometer does not increase cleaning time, we always keep the absorber at the height 32.5 cm. The agreement of the data and the calculation is reasonable; however, analysis of all data yields the accuracy of evaluating the absolute value of spectral cut off not better than 1.0 cm.

**Appendix 2. Evaluation of VUCN detection efficiency, differential storage and escape times**

Evaluation of VUCN detection efficiency in BGS can be found in ref. (13).

Expression for VUCN detection efficiency $\varepsilon(h)$ can be written as:

$$\varepsilon(h) = \frac{\tau_{stor}(h) - \tau_{VUCN}(h)}{\tau_{stor}(h)}. \tag{2.1}$$

Here $\tau_{stor}(h)$ and $\tau_{VUCN}(h)$ are times of storage of neutrons with energy corresponding to the raising height $h$ ($mgh > E_{bound}$) in the Earth's gravitational field in the spectrometer with closed and open exit valve. These values change as a function of sample in the spectrometer and also as a function of temperature. Strictly speaking, this means that the VUCN detection efficiency $\varepsilon(h)$ has to be measured for each sample and temperature.

We used the fact that VUCN cannot be distinguished from UCN in the initial spectrum with energy higher than the gravitational barrier.

To evaluate $\tau_{stor}(h)$ we used the following procedure:

The spectrometer is filled in with neutrons with energy below $h$ (the absorber is set to the height $h$ and stays at this height until the moment of opening of the exit valve, $mgh > E_{bound}$, the exit valve is closed). To shape spectrum, neutrons are stored in the spectrometer for some time $t_1$ after completion of filling (after closing the input valve). Time $t_1$ has to be long enough for removing neutrons with energy higher than $h$ from the spectrometer. After the time interval $t_1$, the exit valve is open and residual neutrons escape from the spectrometer to the detector (neutrons with energy from $E_{bound}$ to $h$). 5 sec before opening the exit valve the absorber is lifted up by 10 cm in order to avoid any influence of residual non-removed energetic neutrons to the data. After completion of neutron counting, the spectrometer is emptied; for this purpose, we open the exit valve and raise the bottom absorber. This procedure is repeated for different $h$ values thus providing $N(h, t_1)$ measurement. $N(h, t_2)$, measured in analogous way, provides the number of UCNs survived in the spectrometer after time $t_2 > t_1$.

These two dependences can be differentiated over energy to give ratio:

$$\frac{dN(h,t_2)/dh}{dN(h,t_1)/dh} = e^{-\frac{\Delta t}{\tau_{stor}(h)}}, \Delta t = t_2 - t_1, \tag{2.2}$$

which describes evolution of the number of UCNs with energy $h$ defined by their loss in spectrometer walls. Thus we can evaluate $\tau_{stor}(h)$. One should note that ratio (2.2) does not contain the efficiency of neutron detection, as it depends on energy but does not depend on time.

In practice, differentiation is replaced by subtraction of results with close absorber heights. Fig. 2.1 gives an example of measurement circles, used to evaluate $N(70\ cm, 70\ sec)$, $N(90\ cm, 70\ sec)$, $N(70\ cm, 220\ sec)$ and $N(90\ cm, 220\ sec)$. In this case expression (3.2) is reduced to:

$$\frac{N(90\ cm, 70\ sec) - N(70\ cm, 70\ sec)}{N(90\ cm, 220\ sec) - N(70\ cm, 220\ sec)} = e^{-\frac{150\ sec}{\tau_{stor}(80\ cm)}}.$$

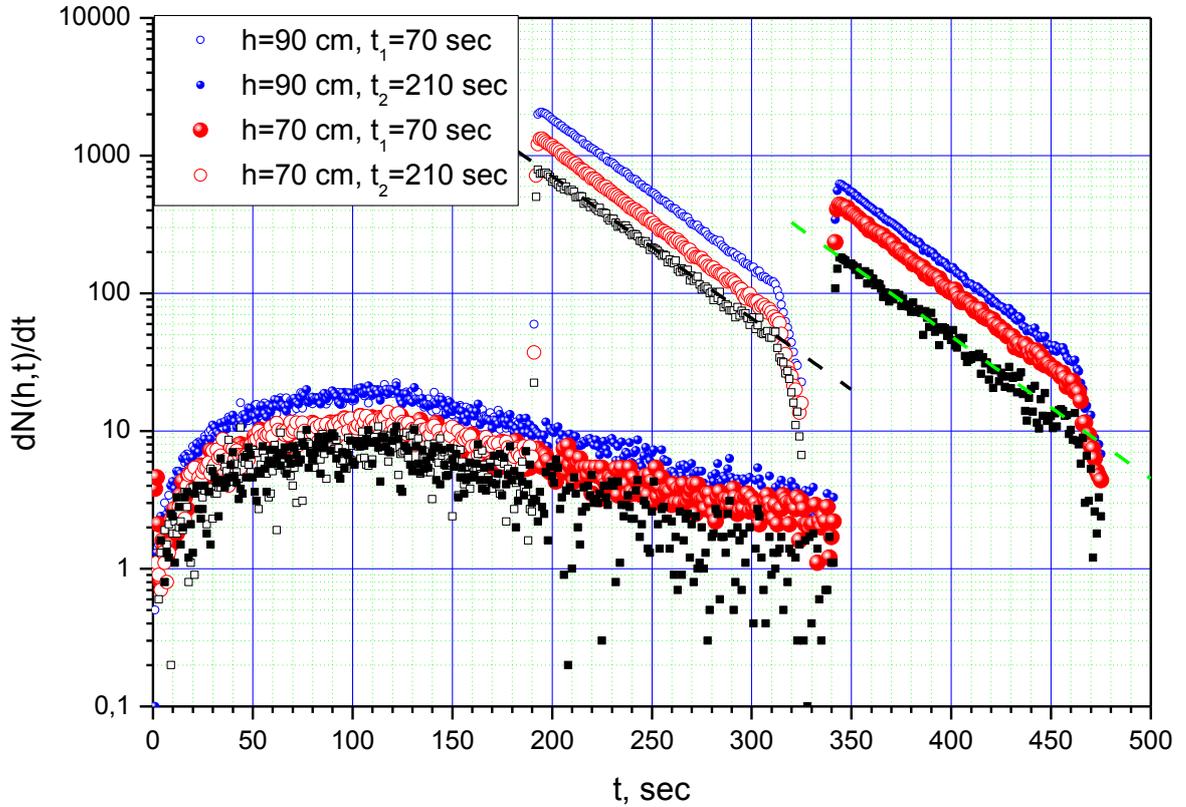

*Fig. 2.1. Neutron count rate in the detector in measurements $\tau_{stor}(h)$ and $\tau_{VUCN}(h)$ as a function of the absorber height; measurements with a thin sample of Fomblin oil.*

*Blue points correspond to measurements with the absorber height equal 90 cm; open points stand for $t_1 = 70\ sec$, closed points for $t_2 = 220\ sec$. Red points show results obtained with the absorber height 70 cm; open points correspond to $t_1 = 70\ sec$, closed points for $t_2 = 220\ sec$. Squares indicate the difference between results of these measurements with different absorber heights, dotted line is a fit of experimental data for evaluating $\tau_{VUCN}(80\ cm)$.*

The time constant $\tau_{VUCN}(h)$ is a characteristic of the rate of decrease of neutrons with energy $h$ in the spectrometer as a result of their loss in the spectrometer walls and in the detector. The rate of this decrease for neutrons with energy from $h_1$ and $h_2$ can be evaluated using per-second difference of detector count rates after opening the exit valve (see Fig. 2.1) at corresponding absorber heights. We fit the obtained curve with exponent function and get $\tau_{VUCN}(h)$.

Results of measurements of $\tau_{stor}(h)$ and $\tau_{VUCN}(h)$ are presented in Fig. 2.2. The efficiency $\varepsilon(h)$ evaluated in accordance with formula (2.1) is shown in Fig. 2.3.

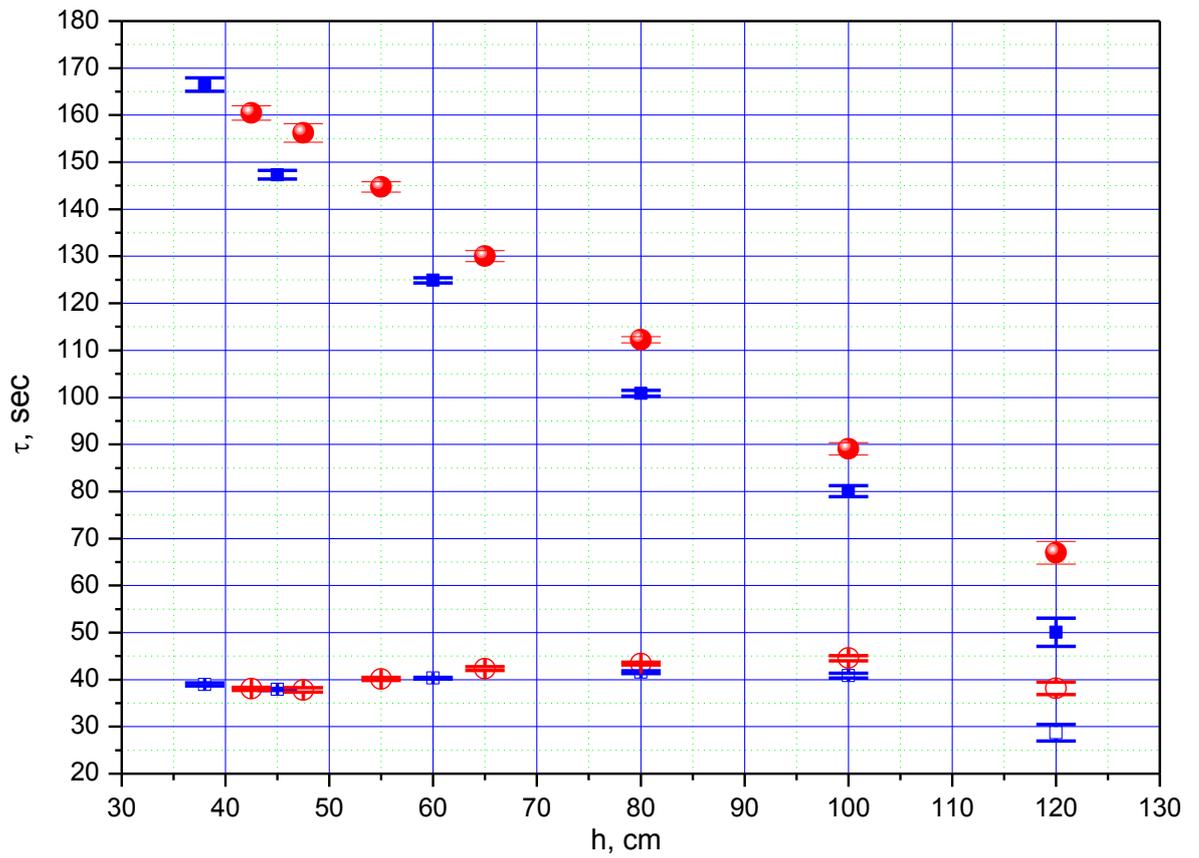

*Fig. 2.2. $\tau_{stor}$ and $\tau_{VUCN}$ are shown as a function of time for empty spectrometer (blue squares) and for the spectrometer with a thin Fomblin oil sample (red circles). Filled in points correspond to $\tau_{stor}(h)$, open points correspond to $\tau_{VUCN}(h)$.*

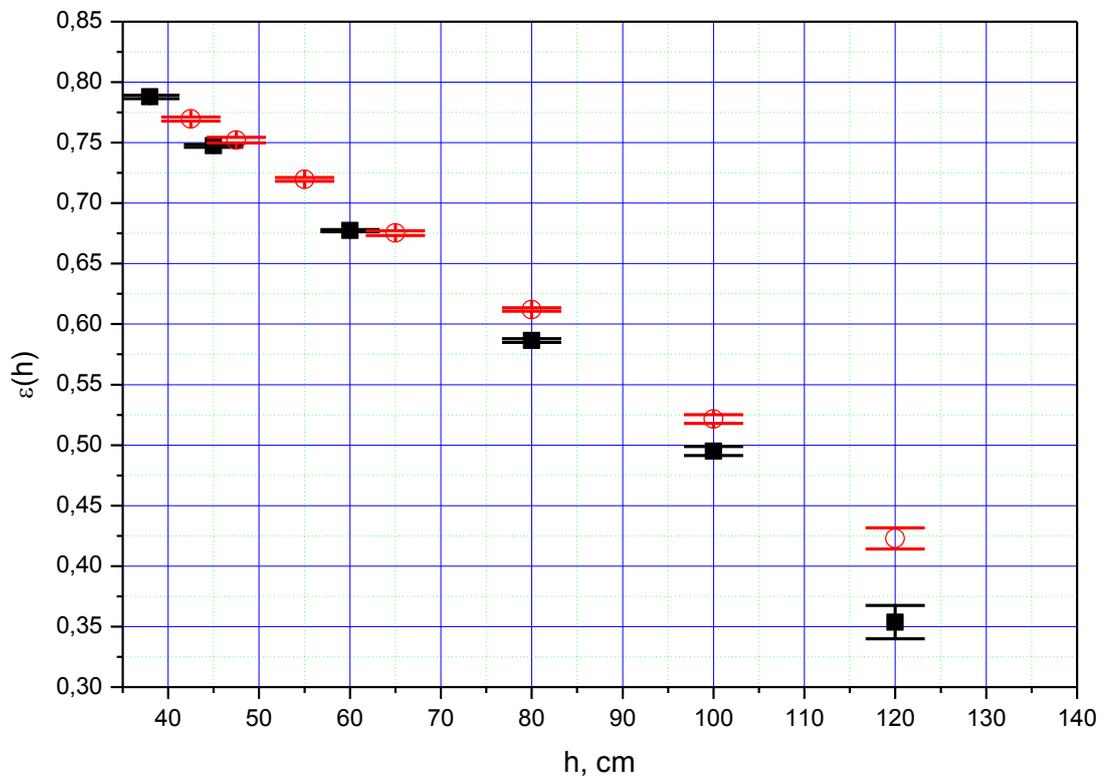

*Fig. 2.3. The efficiency $\varepsilon(h)$ as a function of UCN energy for empty spectrometer with the "lift" moved up (open circles) and for the spectrometer with a thin Fomblin oil sample (black squares).*

Using differential values $\tau_{stor}(h)$ and $\tau_{VUCN}(h)$ we get differential times of emptying the spectrometer for neutrons with energy higher than the gravitational barrier $\tau_{extr}(h)$. These values are related to each other via a simple formula:

$$\frac{1}{\tau_{VUCN}(h)} = \frac{1}{\tau_{stor}(h)} + \frac{1}{\tau_{extr}(h)}. \qquad (1.2)$$

With values $\tau_{stor}(h)$ and $\tau_{VUCN}(h)$ shown in Fig. 2.2 we get values $\tau_{extr}(h)$ shown in Fig. 2.4.

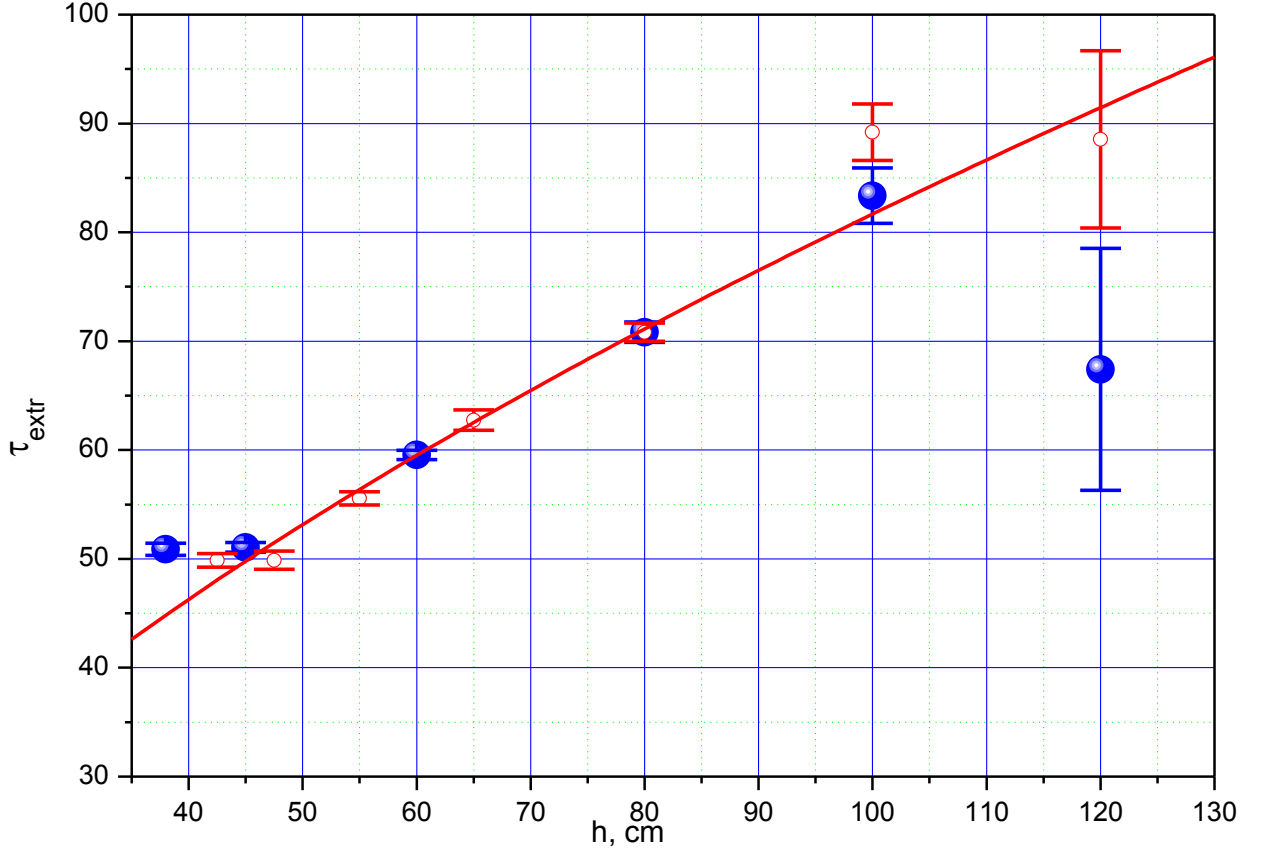

*Fig. 2.4. Emptying time $\tau_{extr}(h)$ as a function of energy for the spectrometer with the "lift" moved up (open circles) and for the spectrometer with a thin Fomblin oil sample (black circles). Curve corresponds to a function $4.7h^{0.62}$ used to fit the data.*

Note that emptying times do not depend on temperature and presence/absence of samples in the spectrometer as they are defined by the setup geometry only. As the main fraction of time in the described procedure of the efficiency evaluation is spent for statistic collection in measurements of $\tau_{extr}(h)$, we do that ones and then, to safe time, measure only differential storage times $\tau_{stor}(h)$ for particular samples; then the efficiency is calculated in accordance with (2.1) and (2.2) as follows:

$$\varepsilon(h) = \frac{\tau_{stor}(h)}{\tau_{stor}(h) + \tau_{extr}(h)}. \qquad (2.3)$$

The data presented in Fig. 2.3 can be fitted with some continuous curve, which, however, does not describe precisely the detection probability for neutrons with energy close to the gravitational barrier (higher than that by 1-2 cm). Large systematic effects for such neutrons are due to the fact that the time of their escape above the gravitational barrier is comparable to or even larger than the times of their emptying to the detector. For neutrons with energy higher than the absorber height, the detection probability is also evaluated using expression (2.3), but their storage time is defined by the probability of their loss in the absorber. Both these dependences (for neutrons with energy close to the barrier and above the absorber) can be calculated; we used a simple model of isotropic UCN gas. Fig. 2.5 illustrates the probability for different absorber heights. These and analogous dependencies are used in model calculations, which are then compared to the experimental data.

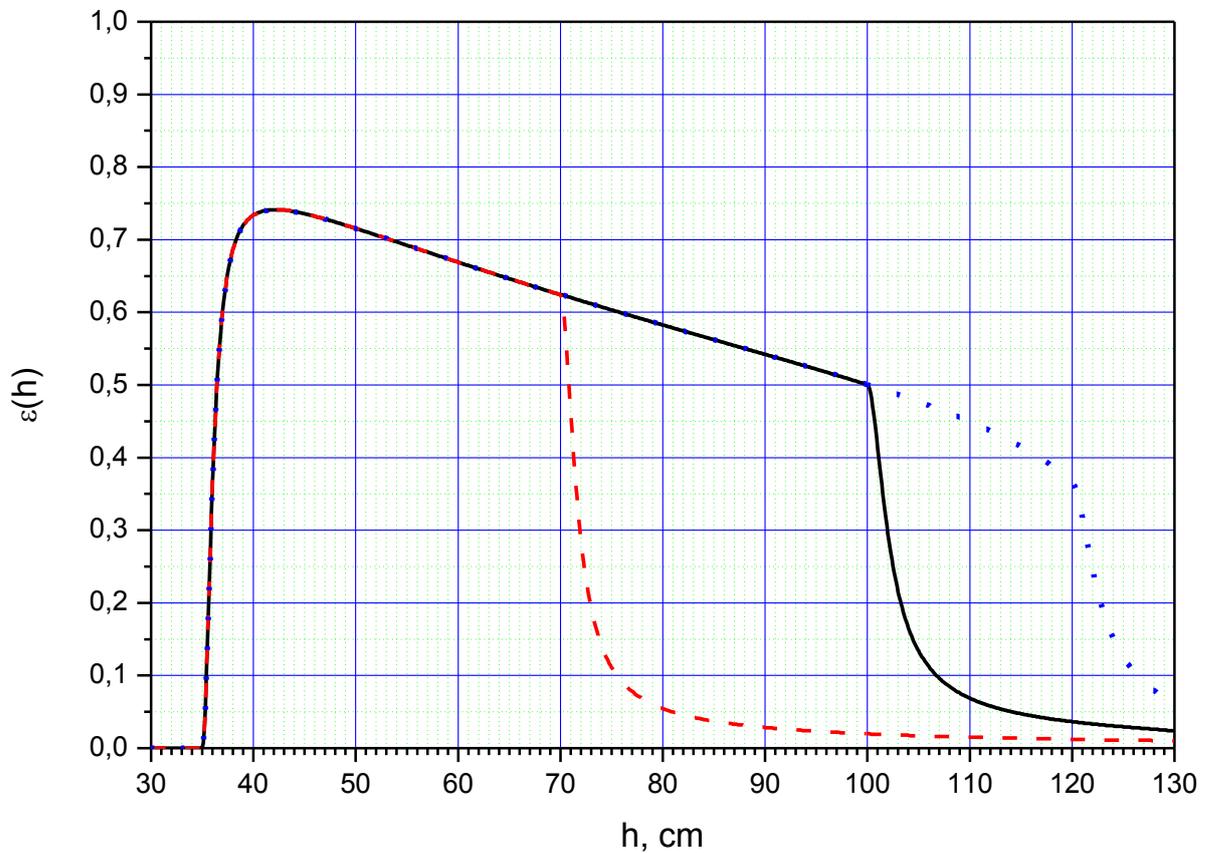

*Fig. 2.5. VUCN detection efficiency as a function of energy in measurements with a thin Fomblin oil sample; for different absorber heights: 70 cm — red dashed line, 100 cm — black solid line, 120 cm — blue dotted line. Central parts of all curves results of fitting the experimental data in Fig. 1.3. Left parts of curves are calculated taking into account time needed to overcome the potential barrier, right parts of curves take into account the effect of absorber to neutron storage times (losses in the absorber are calculated).*

**Appendix 3. UCN spectrum**

Shaping the initial UCN spectrum is an important part of the measuring procedure. The main purpose is to avoid false effects associated with neutrons survived after filling the spectrometer with initial energy higher than the gravitational barrier height. Besides, the knowledge of initial UCN spectrum in the spectrometer is needed for correct comparison of the data with theoretical calculations. We aim at shaping narrow initial spectrum; its width should be sufficient for statistics collection; its upper cut off should be close to the gravitational barrier; false effects should not be observed.

UCN spectrum is evaluated by measuring the number of neutrons survived in the spectrometer after cleaning. Measuring procedure consists of the following stages: filling (60 sec), cleaning (45 sec), storage (5 sec) and emptying (150 sec). During filling and cleaning the absorber is installed to the height $H_{abs} < H_{bound}$, during storage and emptying it is installed to the height $H_{abs} + 5\ cm$. During emptying the monitor valve is open and neutrons from the internal part of the spectrometer storage volume are counted in the monitor detector. The measurement is repeated with different values $H_{abs}$, thus providing the number of neutrons in the spectrometer as a function of the absorber height. Fig. 3.1 shows results of such measurement with a thin Fomblin oil sample with the area 0.74 m².

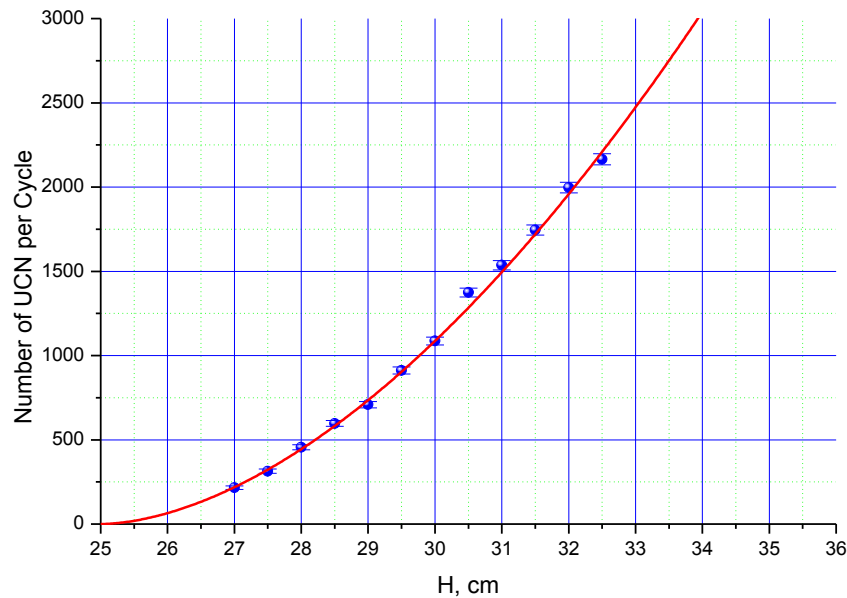

*Fig. 3.1. The number of counted UCN as a function of the absorber height. Solid line is a result of fitting with a function $\cdot (H - H_0)^k$.*

Typical count rate in the monitor detector is presented in Fig. 3.2. From measured evolution of the monitor count rate after opening of the monitor valve we can evaluate the characteristic time of emptying neutrons from the storage volume $\tau_{UCN}$, which is related to the storage time $\tau_{st}$ and the time of emptying to the monitor $\tau_m$:

$$\frac{1}{\tau_{UCN}} = \frac{1}{\tau_m} + \frac{1}{\tau_{st}}. \tag{2.1}$$

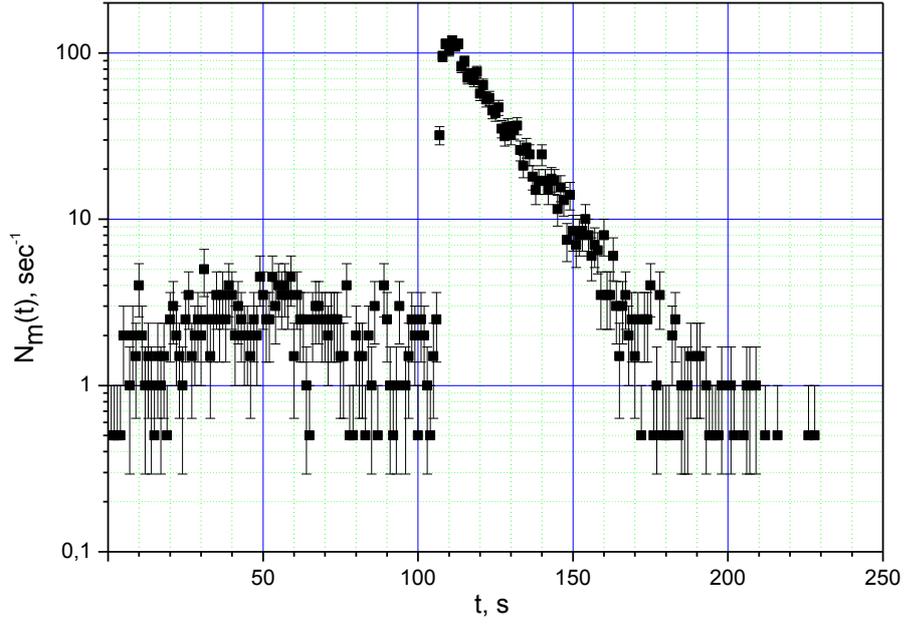

*Fig. 3.2. Monitor count rate as a function of time in measurements of the initial UCN spectrum with the absorber height $H_{abs} = 32.5\ cm$ with a Fomblin oil sample with the surface area 0.74 $m^2$.*

During emptying the absorber is lifted up to avoid eventual effects of neutrons, which had not been cleaned, to the measured curve. As the UCN spectrum is quite narrow, one can assume that UCN storage time $\tau_{st}$ is approximately the same and can be extracted from the time evolution of the monitor count rate in main measurements (Fig. 2). Time constants evaluated in such a way from the data are: $\tau_{st} = 146 \pm 2\ sec$ and $\tau_{UCN} = 15.0 \pm 1.6\ sec$. Thus, according to expression (2.1) we get $\tau_m = 16.7 \pm 1.8\ sec$.

Curve in Fig. 3.1 is well described by function $F_{int}(h) = 65 \cdot (h - 25)^{1.75}$, where the height $h$ is measured in cm.

As, due to the small width of the spectrum, we assumed that $\tau_m$ and $\tau_{st}$ are about the same for all neutrons in the spectrometer, then $F_{int}(h)$ is proportional to the integral UCN spectrum. Accordingly, the differential neutron spectrum is proportional to $\Phi_{exp}(h) = 65 \cdot 1.75 \cdot (h - 25)^{0.75}$, and the number of UCNs in the spectrometer at the beginning of cleaning can be described as follows:

$$\Phi_0(h) = \Phi_{exp}(h) \cdot \frac{\tau_{st}}{\tau_{st} - \tau_m} \cdot e^{-\frac{\Delta t_{clean}}{\tau_{st}}}. \tag{3.2}$$

Here second factor takes into account that emptying neutrons to the monitor detector takes time and therefore a fraction of neutrons is lost in the spectrometer during this time. Third factor takes into account loss of neutrons in the spectrometer walls and in the sample during cleaning. It is natural to assume that function (3.2) describes UCN spectrum also in the range of $h$ slightly above 32.5 cm (maximum height of the absorber in measurements of UCN spectrum). Then UCN spectrum in the spectrometer for a given $H_{abs}$ at the end of cleaning can be described as follows:

$$\Phi(h) = \Phi_0(h) \cdot \left( e^{-\frac{\Delta t_{clean}}{\tau_{st}}} \cdot \theta(H_{abs} - h) + e^{-\Delta t_{clean}\left(\frac{1}{\tau_{st}} + \frac{1}{\tau_{cl}(h,H_{abs})}\right)} \cdot \theta(h - H_{abs}) \right), \quad (3.3)$$

where $\theta(x)$ is Heaviside function, and $\tau_{cl}(h, H_{abs})$ is defined by expression (1.4).

Fig. 3.4 shows UCN spectrum in accordance with expression (3.3) for the absorber height $H_{abs} = 32.5\ cm$ and calculated loss of neutrons in the absorber.

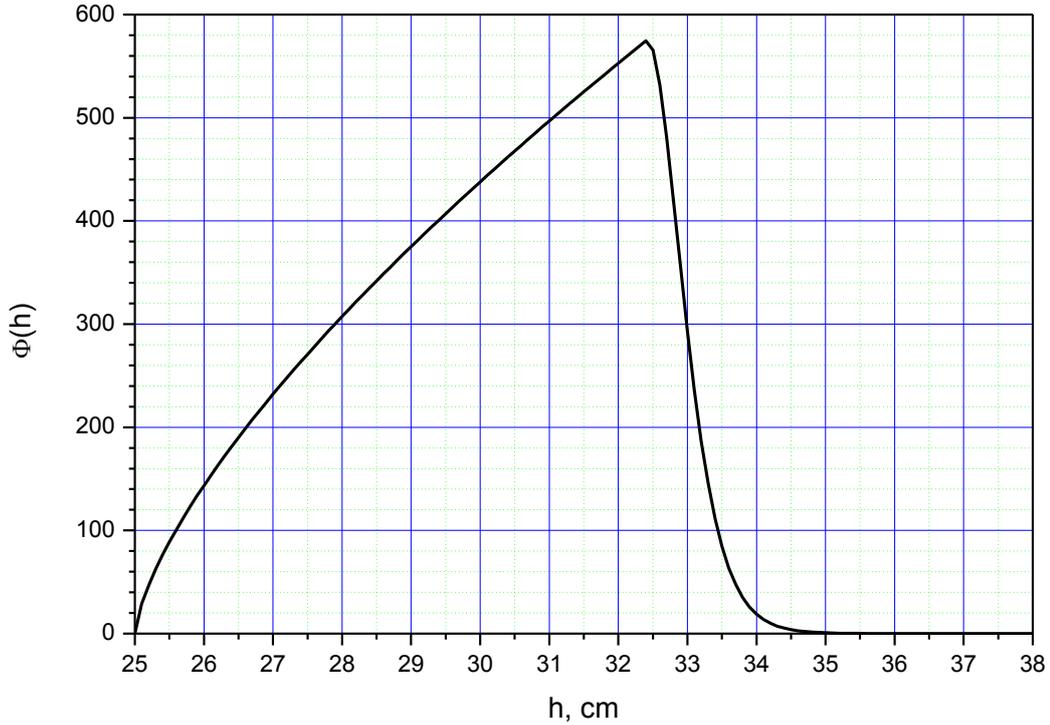

*Fig. 3.4. UCN spectrum in the spectrometer with a thin Fomblin oil sample with the area 0.74 m² at the end of spectrum cleaning.*

**References**
1. **Nesvizhevsky, V.V., et al.** *Europ. Phys. J. App. Phys.* 6 (1999) 151.
2. **Nesvizhevsky, V.V., et al.** *Phys. At. Nucl.* 62 (1999) 776.
3. **Ignatovich V.K.** *The Physics of Ultracold Neutrons.* s.l. : Clarendon Press, 1990.
4. **Barabanov, A.L., et al.** *Europ. Phys. J. B.* 15 (2000) 59.
5. **Barabanov, A.L., et al.** *Eur. Phys. J. A.* 27 (2006) 105.
6. **Strelkov, A.V, et al.** *Nucl. Instr. Meth. A.* 440 (2000) 695.
7. **Lychagin, E.V., et al.** *Phys. At. Nucl.* 63 (2000) 548.
8. **Lychagin, E.V., et al.** *Phys. At. Nucl.* 65 (2002) 1995.
9. **Steyerl, A., et al.** *Eur. Phys. J. B.* 28 (2002) 299.
10. **Serebrov, A.P., et al.** *Phys. Lett. A.* 309 (2003) 218.
11. **Bondarenko, L., et al.** *JETP Lett.* 68 (1998) 691.
12. **Bondarenko, L.N., et al.** *Phys. At. Nucl.* 65 (2002) 11.
13. **Kartashev, D.G., et al.** *Int. J. Nanoscience.* 6 (2007) 501.
14. **Canaguier-Durand, A., et al.** *Phys. Rev. A.* 83 (2011) 032508.
15. **Nesvizhevsky, V.V., et al.** *New J. Phys.* 14 (2012) 093053.
16. **Nesvizhevsky, V.V., et al.** *Crystal. Rep.* 58 (2013) 743.

17. **Nesvizhevsky, V.V.** *Phys. At. Nucl.* 65 (2002) 400.
18. **Pokotilovski, Y.N.** *Phys. Lett. A.* 255 (1999) 173.
19. **Lamoreaux, S.K., et al.** *Phys. Rev. C.* 66 (2002) 044309.
20. **Olive, K.A., et al (Particle Data Group).** *Chin. Phys. C.* 38 (2014) 090001.
21. **Steyerl, A., et al.** *Phys. Rev. C.* 85 (2012) 065503.
22. **Pokotilovski, Yu.N., et al.** *Phys. B.* 403 (2008) 1942.
23. **http://www.solvay.com/en/binaries/Fomblin-PFPE-Lubes-for-Vacuum-Applications_EN-220533.pdf. [Online]**
24. **Mampe, W., et al.** *Phys. Rev. Lett.* 63 (1989) 593.
25. **Mampe, W., et al.** *JETP Lett.* 57 (1993) 82.
26. **Arzumanov, S., et al.** *Phys. Lett. B.* 483 (2000) 15.
27. **Pichlmaier, A., et al.** *Nucl. Instr. Meth. A.* 440 (2000) 517.
28. **Pichlmaier, A., et al.** *Phys. Lett. B.* 693 (2010) 221.
29. **Arzumanov, S., et al.** *Phys. Lett. B.* 745 (2015) 79.
30. **http://alter-b.ru/upload/files/fom_vac.pdf. [Online]**
31. **http://www.solvayplastics.com/sites/solvayplastics/EN/Solvay%20Plastics%20Literature/BR_Fomblin_Vacuum.pdf. [Online]**
32. **http://www.solvayplastics.com/sites/solvayplastics/EN/specialty_polymers/Fluorinated_Fluids/Pages/Fomblin_Lubricants_PFPE.aspx. [Online]**
33. **Arzumanov, S.S., et al.** s.l. : ISINN-10, 2002. p. 356.
34. **https://www.goodfellowusa.com/catalog/GFCat4J.php?ewd_token=cUvdNgdBjp9xJ8jPra0kd5etVL3Y09&n=b7kICdbJZT7TgVgzBqIxVAys1xI89s. [Online]**
35. **Vertes, M.** *Vacuum.* Vol. 44 (1993) 769.
36. **Nesvizhevsky, V.V.** 1993. ILL Preprint 93VVN02S.